\documentclass[10pt]{article}
\usepackage{spconf}

\usepackage{verbatim}
\usepackage{amsfonts,amsmath,mathrsfs,amssymb,amsbsy}
\usepackage[final]{graphicx}
\usepackage{times,cite}
\usepackage{subfig}

\long\def\symbolfootnote[#1]#2{\begingroup%
\def\thefootnote{\fnsymbol{footnote}}\footnote[#1]{#2}\endgroup}

\usepackage{verbatim}
\usepackage[]{float,latexsym}
\usepackage{amsfonts,amsmath,mathrsfs,amssymb,amsbsy}
\usepackage{url}

\newtheorem{theorem}{Theorem}[section]

\newcommand{\Prob}{\mathsf{P}}
\newcommand{\Expect}{\mathsf{E}}

\usepackage{color}
\definecolor{lightblue}{rgb}{.7, .8, 1}
\definecolor{lightgreen}{rgb}{.6, 1, .6}
\usepackage{color}
\definecolor{brown}{rgb}{1,0.38,0.03}

\definecolor{OliveGreen}{rgb}{.2,0.6,0.2}
\definecolor{BrickRed}{rgb}{.7,0.2,0.2}

\newcommand{\ignore}[1]{} 

\long\def\symbolfootnote[#1]#2{\begingroup%
\def\thefootnote{\fnsymbol{footnote}}\footnote[#1]{#2}\endgroup}

\DeclareMathOperator*{\argmin}{arg\,min}

\newcommand{\bsp}{\begin{split}}

\newcommand{\esp}{\end{split}}

\begin{document}

\sloppy
\ninept

\title{Classification of Local Field Potentials using Gaussian Sequence Model}



\name{Taposh Banerjee$^{\star}$ \; John Choi$^{\dagger}$ \; Bijan Pesaran$^{\dagger}$ \; Demba Ba$^{\star}$ \; and \; Vahid Tarokh$^{\star}$ \thanks{The work at both Harvard and NYU is supported by the Army Research Office MURI under Contract Number W911NF-16-1-0368.}}

\address{$^{\star}$ School of Engineering and Applied Sciences, Harvard University\\
    $^{\dagger}$ Center for Neural Science, New York University}
\maketitle

\begin{abstract}
A problem of classification of local field potentials (LFPs), recorded from the prefrontal cortex of a macaque monkey, is considered. 
An adult macaque monkey is trained to perform a memory-based saccade. The objective is to decode the eye movement goals 
from the LFP collected during a memory period. The LFP classification problem is modeled as that of 
classification of smooth functions embedded in Gaussian noise. It is then argued that using minimax function estimators as features 
would lead to consistent LFP classifiers. The theory of Gaussian sequence models allows us to represent minimax estimators as finite dimensional objects. The LFP classifier resulting from this mathematical endeavor is a spectrum based technique, 
where Fourier series coefficients of the LFP data, followed by appropriate shrinkage and thresholding, are used as features in a linear discriminant classifier. The classifier is then applied to the LFP data to achieve high decoding accuracy. The function classification 
approach taken in the paper also provides a systematic justification for using Fourier series, with shrinkage and thresholding, as features for the problem, as opposed to using the power spectrum. It also suggests that phase information is crucial to the decision making. 
\end{abstract}


\section{Introduction}
One of the most important challenges in the development of effective brain-computer interfaces is the decoding of brain signals. 
Although we are still very far from obtaining a complete understanding of the brain, yet important contributions have been made. 
For example, results are reported in the literature where local field potential (LFP) data or spike data are collected from animal species, while the animal is performing a task. The LFP or spike data are then used to make various inferences regarding the task. See for example \cite{rao2013brain} and \cite{markowitz2011optimizing}.

Spectrum-based techniques are popular in the literature. Fourier series, wavelet transform or power spectrum of the recorded signals 
are often used as features in the machine learning algorithms used for inference. However, the choice for such spectrum based techniques is either ad-hoc or is not well motivated. 

In this paper, we propose a mathematical framework for classification of LFP signals into one of many classes or categories. The framework is of Gaussian sequence model for nonparametric estimation; see \cite{Johnstone2015Book} and \cite{tsybakov2009introduction}. The LFP data is collected from the prefrontal cortex of a macaque monkey, while the monkey is 
performing a memory-based task. In this task the monkey saccades to one of eight possible target locations. 
The goal is to predict the target location using the LFP signal. More details about the task are provided in Section~\ref{sec:MemExpt} below. 

The LFP signal is modeled as a smooth function corrupted by Gaussian noise. The smooth signal is part of the LFP relevant to the classification problem. The classification problem is then formulated as the problem of classifying 
the smooth functions into one of finitely many classes (Section~\ref{sec:NPFramework}). We provide mathematical arguments to justify that using minimax function estimator for the smooth signal as a feature leads to a consistent classifier (Section~\ref{sec:ClassUsingMinimaxEst}). The theory of Gaussian sequence models allows us to obtain minimax estimators and also to represent them as finite dimensional vectors (Section~\ref{sec:GaussSeqMod}).
We propose two classification algorithms based on the Gaussian sequence model framework (Section~\ref{sec:ClassifyGSM}). The classification algorithms involve computing the Fourier series coefficients of the LFP data and then using them as features after appropriate shrinking and thresholding. The classifiers are then applied to the LFP data collected from the monkey and provide
high decoding accuracy of up to $88$\% (Section~\ref{sec:Numerical}). 

The Gaussian sequence model approach used in this paper leads to the following insights:
\begin{enumerate}
\item A systematic modeling of the classification problem leads to Fourier series coefficients, with shrinkage and thresholding, to be used 
as features for the problem (as opposed to the use of power spectrum for example). 
\item Numerical results in Section~\ref{sec:Numerical} show that using the absolute value of the Fourier series coefficients as features results in a drop of $15$\% in accuracy. This suggests that phase information is crucial to the classification problem studied in this paper. 
\end{enumerate}

\section{A Memory-Based Saccade Experiment}\label{sec:MemExpt}
\begin{figure}[t]
	\center
	\includegraphics[width=6cm, height=5cm]{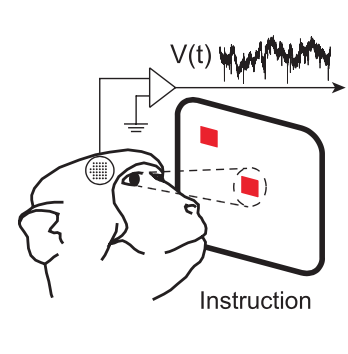}
	\includegraphics[width=6cm, height=4cm]{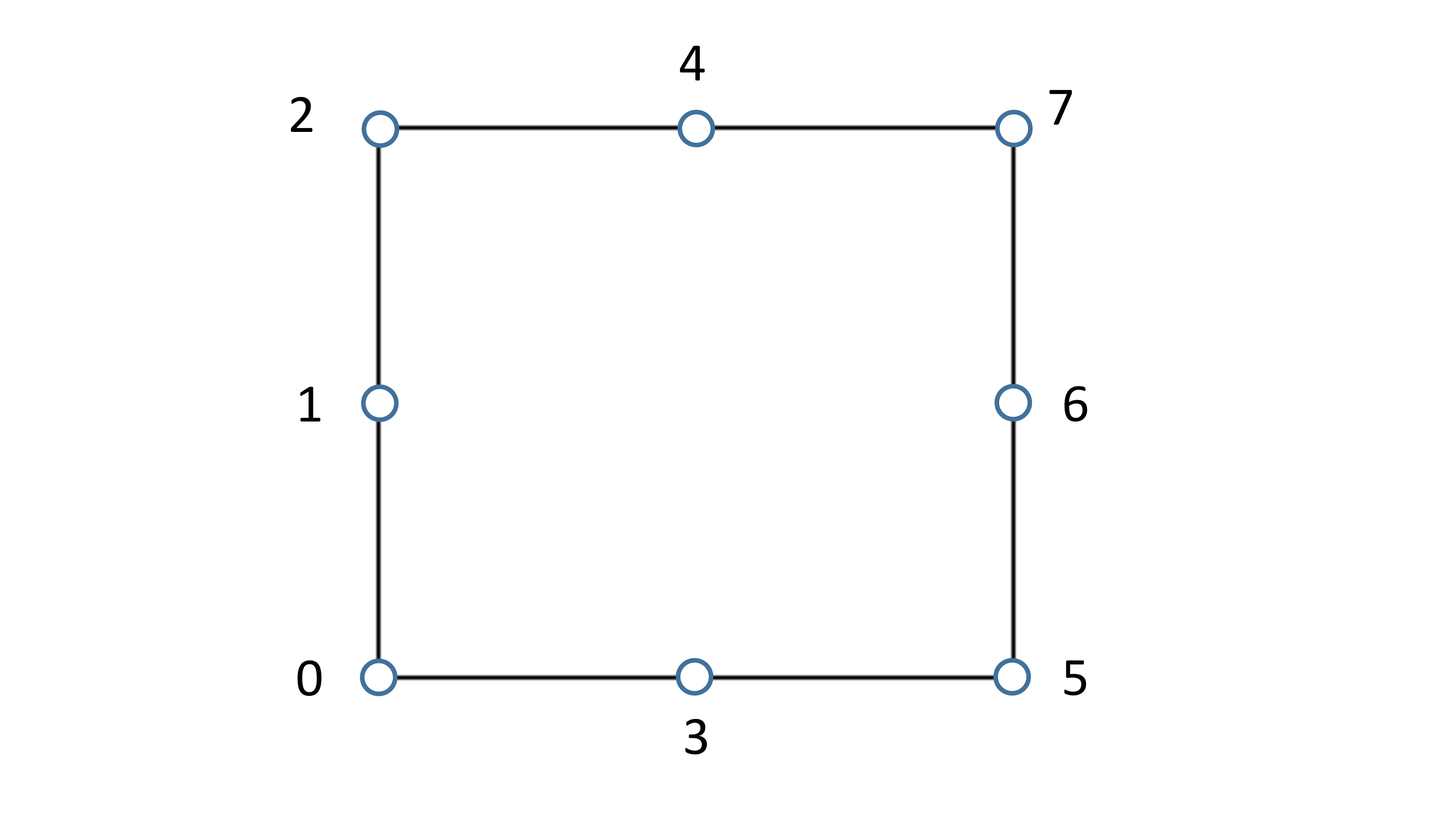}
	\caption{Memory task and target locations from \cite{markowitz2011optimizing}.}
		\label{fig:MonkeyExprt}
\end{figure}

An adult macaque monkey is trained to perform a memory-based saccade. A single trial starts with the monkey looking at the center of a square target board; see Fig.~\ref{fig:MonkeyExprt}. The center is illuminated with an LED light, and the monkey is trained to fixate at the center light at the beginning of the trial. After a small delay, a target light at one of eight corners of the target board (four vertices and four side centers) is switched on. The location of the target is randomly chosen across trials. After another short delay, the target light is switched off. The switching
off of the target light marks the beginning of a memory period. After some time the center light is switched off. The latter provides a
cue to the monkey to saccade to the location of the target light, the location where the target was switched on before switching off. 
The cue also marks the end of the memory period. 
If the monkey correctly saccades to the target chosen for the trial, i.e., if the monkey saccades to within a small area around the target, then the trial is declared a success. Only data collected from successful trials is utilized in this paper. 

LFP and spike data are collected from electrodes embedded in the cortex of the monkey throughout the experiment. 
The data is collected using a 32 electrode array resulting in 32 parallel streams of LFP data. 
The data used in this paper has been first collected and analyzed in \cite{markowitz2011optimizing}. 
Further details about sampling rates, timing information and the post-processing done on the data, can be found in \cite{markowitz2011optimizing}. 

 It is hypothesized that different target locations would correspond to different LFP and spike rate patterns. The objective is thus to detect 
 or predict one of eight target locations based on the LFP data after the target is switched off and before the cue, i.e., the LFP data collected during 
 the memory period. 

\section{A Nonparametric Regression Framework for LFP Data}\label{sec:NPFramework}
Let $Y_\ell$, for $\ell \in \{0,1, \cdots, N-1\}$, represent the sampled version of the LFP waveform collected from one of the 32 channels during the memory period. We model the LFP data in a non-parametric regression framework:
\begin{equation}\label{eq:RegressLFPMod}
Y_\ell = f( \ell / N ) +  Z_\ell, \quad \ell \in \{0, \cdots, N-1\}.
\end{equation}
Here, $f: [0,1] \to \mathbb{R}$ is a smooth function, and represent the part of the LFP waveform that is relevant to the classification problem. 
Thus, $f(t)$ is the signal that contains information about where the monkey will saccade, out of eight possible target locations, 
after the cue. 
Also, $Z_\ell \stackrel{\text{i.i.d.}}{\sim}  \mathcal{N}(0,1)$, and correspond to neural firing data either not pertinent to the memory experiment itself, or corresponds to the experiment, but not relevant to the classification task. We model this impertinent data as noise, and treat $f(t)$ as the signal. We assume that $f \in \mathcal{F}$, where $\mathcal{F}$ is a class of smooth functions in $L_2[0,1]$. Precise assumptions about $\mathcal{F}$ will be made later; see Section~\ref{sec:ClassifyGSM} below.
In Fig.~\ref{fig:sampleLFP} we show a sample LFP waveform and its reconstruction using first $10$ Fourier coefficients. The figure
shows that a nonparametric regression framework to model the LFP data is appropriate. 

\begin{figure}[t]
	\center
	\includegraphics[width=9cm, height=6cm]{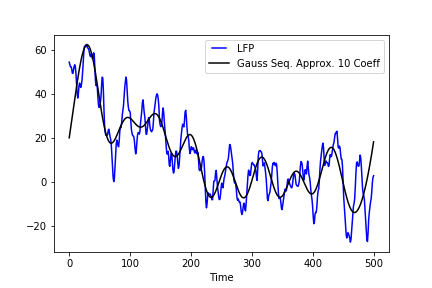}
	\caption{Sample LFP waveform and its reconstruction using the first 10 Fourier series coefficients. }
		\label{fig:sampleLFP}
\end{figure}

We hypothesize that the signal $f$ will be different for all the eight different target locations. In fact, even if the monkey saccades to 
the same location over different trials, the signal $f$ cannot be expected to be exactly the same. Thus, we assume that for target location $k$, $k \in \{1, 2, \cdots, 8\}$, $f \in \mathcal{F}_k$, where $\mathcal{F}_k$ is a subset of $\mathcal{F}$, $\mathcal{F}_k \cap \mathcal{F}_j = \emptyset$, for $j \neq k$. 
Thus, our classification problem can be stated as
\begin{equation}\label{eq:classProb}
\begin{split}
H_k: \quad f \in \mathcal{F}_k, & \quad k \in \{1, 2, \cdots, 8\}, \\
& \mathcal{F}_k \subset \mathcal{F}, \quad \mathcal{F}_k \cap \mathcal{F}_j = \emptyset, \mbox{ for } j \neq k.
\end{split}
\end{equation}      
Our objective is to find a classifier that maps the observed time series $\{Y_\ell\}$ into one of the eight possible target locations. 

\section{Classification Using Minimax Function Estimators}\label{sec:ClassUsingMinimaxEst}
Let $\hat{f}_N$ be an estimator for function $f$ based on $(Y_0, \cdots, Y_{N-1})$ such that its maximum estimation error over 
the function class $\mathcal{F}$ goes to zero:
\begin{equation}\label{eq:fNerrconZero}
\sup_{f \in \mathcal{F}}\; \Expect_f[ \| \hat{f}_N - f \|_2^2] \to 0, \; \mbox{ as } N\to \infty. 
\end{equation}
Here $\Expect_f$ is the expectation with respect to the law of $(Y_0, \cdots, Y_{N-1})$ when $f$ is the true function, and 
$\|f - g\|_2 = \int_0^1 (f(t) - g(t))^2 dt$ is the $L_2$ function norm.  

Let $\delta_{\text{md}}$ be the minimum distance decoder:
\begin{equation}\label{eq:minDistDec}
\delta_{\text{md}}(f) = \argmin_{k \leq  8} \| f - \mathcal{F}_k\|_2 := \argmin_{k \leq  8} \inf_{g \in \mathcal{F}_k}\| f - g\|_2.
\end{equation}
Of course, to implement this decoder we need to know the function classes $\{\mathcal{F}_k\}_{k=1}^8$.

We now show that if the function classes $\{\mathcal{F}_k\}_{k=1}^8$ are known, then $\delta_{\text{md}}(\hat{f}_N)$, 
the minimum distance decoder using $\hat{f}_N$, is consistent. In the rest of this section, we use $\hat{\delta}$ to denote $\delta_{\text{md}}(\hat{f}_N)$ for brevity. 

Let $P_e$ be the worst case classification error defined by
\[
P_e = \max_{k \in \{1, \cdots, 8\}} \; \sup_{f \in \mathcal{F}_k} \; \Prob_f(\hat{\delta} \neq k).
\]
Also, let for $s > 0$, the classes are separated by a distance of at least $2 s$, i.e., 
\begin{equation}\label{eq:2sseperation}
\min_{k \neq  j} \| \mathcal{F}_j - \mathcal{F}_k\|_2 := \min_{k \neq  j} \inf_{f \in \mathcal{F}_j, \; g \in \mathcal{F}_k}\| f - g\|_2 \; > \; 2s.
\end{equation}
Then, for $f \in \mathcal{F}_k$ and $\hat{\delta} = \delta_{\text{md}}(\hat{f}_N)$,
\begin{equation}\label{eq:consitencyOfConderr}
\begin{split}
\Prob_f(\hat{\delta} \neq k) & \; \leq \; \Prob_f ( \|\hat{f}_N  - \mathcal{F}_k\|_2  > s)\\
                                           & \; \leq \; \Prob_f ( \|\hat{f}_N  - f\|_2  > s) \\
                                           & \; \leq \; \frac{1}{s^2}\; \Expect_f [  \|\hat{f}_N  - f\|_2^2]\\
                                           & \; \leq \; \frac{1}{s^2}\; \sup_{g \in \mathcal{F}} \; \Expect_g [  \|\hat{f}_N  - g\|_2^2] \to 0, \quad \mbox{ as } N \to \infty,
\end{split}
\end{equation}
where the limit to zero follows from \eqref{eq:fNerrconZero}.
In fact, since the right most term of \eqref{eq:consitencyOfConderr} is not a function of $k$ and $f$, we have 
\begin{equation}\label{eq:consistecyOfMaxErr}
\begin{split}
P_e = \max_{k \in \{1, \cdots, 8\}} \; & \sup_{f \in \mathcal{F}_k}  \; \Prob_f(\hat{\delta} \neq k) \leq \\
&\frac{1}{s^2}\; \sup_{g \in \mathcal{F}} \;\Expect_g [  \|\hat{f}_N  - g\|_2^2] \to 0, \quad \mbox{ as } N \to \infty.
\end{split}
\end{equation}

Thus, the worst case classification error goes to zero if we use a function estimator with property \eqref{eq:fNerrconZero}. If we choose the minimax estimator $\hat{f}^*_N$, i.e., $\hat{f}^*_N$ satisfies 
\begin{equation}\label{eq:minimaxErrConZero}
\inf_{\hat{f}} \sup_{f \in \mathcal{F}}\; \Expect_f[ \| \hat{f} - f \|_2^2] \sim \sup_{f \in \mathcal{F}}\; \Expect_f[ \| \hat{f}^*_N - f \|_2^2], \; \mbox{ as } N\to \infty,
\end{equation}
then we will have the fastest rate of convergence to zero (fastest among all classifiers of this type) for the misclassification error in \eqref{eq:consistecyOfMaxErr}. Note further that the above consistency result is valid for any choice of function classes $\{\mathcal{F}_k\}$, as long as the classes are well separated, as in \eqref{eq:2sseperation}. This fact demonstrates a robust nature of the minimax estimator based classifier $\delta_{\text{md}}(\hat{f}^*_N)$

\medskip

In practice, the function classes $\{\mathcal{F}_k\}$ are not known, and one has to learn them from samples of data. 
In machine learning, a feature is often extracted from the observations, and the feature is then used to train a classifier. 
The above discussion on minimax estimators suggests that if the class boundaries can be reliably learned, then 
a classifier based on the minimax estimator and the minimum distance decoder is consistent and robust.
This motivates the use of the minimax estimator itself as a feature. That is, given the data, use the minimax estimators 
to learn the class boundaries.  

A typical function estimator is an infinite-dimensional object, and hence it is a complicated object to choose as a feature. However, 
the theory of Gaussian sequence models, to be discussed next, allows us to map the minimax estimator to a finite-dimensional vector (under certain assumptions on $\mathcal{F}$). We use the finite-dimensional representation of the minimax estimator to learn the 
class boundaries and train a classifier. 

\section{Gaussian Sequence Model}\label{sec:GaussSeqMod}
In this section, we provide a brief review of nonparametric or function estimation (Section \ref{sec:NPEst}) and its connection with the Gaussian sequence model (Section~\ref{sec:GaussSeqModel}).
We also state the Pinsker's theorem, that is the fundamental result on minimax estimation on compact ellipsoids (Section~\ref{sec:PinskersThm}).
Finally, we discuss the blockwise James-Stein estimator, that has adaptive minimaxity properties (Section~\ref{sec:BJS}). 
In Section~\ref{sec:ClassifyGSM}, we discuss the applications of these results to classification in neural data.

\subsection{Nonparametric Estimation}\label{sec:NPEst}
Consider the problem of estimating a function $f: [0,1] \to \mathbb{R}$ from its noisy observation $Y: [0,1] \to \mathbb{R}$, where $Y(t)$
now does not necessarily represent an LFP signal, but is a more abstract object. 
The precise observation model(s) will be provided below. Let $\hat{f}: [0,1] \to \mathbb{R}$ be our estimate. If the function and its estimate are in $L_2([0,1])$, 
then one can use $\int_0^1 (f(t) - \hat{f}(t))^2 dt$ as an estimate of the error. This error estimate is random due to the randomness of the noise. Thus, a more appropriate error criterion is
\begin{equation}\label{eq:MSE}
\text{MSE}(\hat{f}) = \Expect_f \left[\int_0^1 (f(t) - \hat{f}(Y; t))^2 \; dt \right],
\end{equation}
where the dependence of estimate $\hat{f}$ on the observation $Y$ is made explicit, and also we assume that the expectation is well defined for the noise model in question. 

Two popular models for the observation $Y$ are as follows.

\medskip
\textit{White Noise Model}: Here the observation Y is given by the stochastic differential equation \cite{rogersWilliams}
\begin{equation}\label{eq:WhiteNoiseMod}
Y(t) = \int_0^t f(s) \; ds + \epsilon \; W_t, \quad t\in [0,1],
\end{equation}
where $W$ is a standard Brownian motion, and $\epsilon > 0$. 

\medskip
\textit{Regression Model}: Here the observation $Y$ is a discrete stochastic process corresponding to samples of the function $f$ at discrete points, 
\begin{equation}\label{eq:RegressionMod}
Y_k = f( k / N ) +  Z_k, \quad k \in \{0, \cdots, N-1\},
\end{equation}
where $Z_k \stackrel{\text{i.i.d.}}{\sim} \mathcal{N}(0,1)$, and $N$ is the number of observations. Although we only use the regression model for classification of LFP in this paper, the discussion of white noise model is important as it provided an exact map to a Gaussian sequence model (see Section~\ref{sec:GaussSeqModel} below for precise statements). 

\medskip
In the theory of nonparametric estimation \cite{tsybakov2009introduction}, minimax optimality of estimators like kernel estimators, local polynomial estimators, projection estimators, etc. is investigated. It is shown that if the function $f$ is smooth enough, then carefully designed versions of these type of estimators are asymptotitcally optimal in a minimax sense, with $N \to \infty$ in the regression model \eqref{eq:RegressionMod}, and $\epsilon \to 0$ in the white noise model \eqref{eq:WhiteNoiseMod}. 

\subsection{Gaussian Sequence Model}\label{sec:GaussSeqModel}
One of the most remarkable observations in nonparametric estimation theory is that the estimation in the above two models, the white noise model \eqref{eq:WhiteNoiseMod} and the regression model \eqref{eq:RegressionMod}, are fundamentally equivalent. Moreover, these two are also equivalent to many other popular estimation problems, including density estimation and power spectrum estimation.
The common thread that ties them together is the Gaussian sequence model \cite{Johnstone2015Book}.  

A Gaussian sequence model is described by an observation sequence $\{y_k\}$:
\begin{equation}\label{eq:GaussSeqMod}
y_k = \theta_k + \epsilon \; z_k, \quad k \in \mathbb{N}.
\end{equation} 
The sequence $\{\theta_k\}$ is the unknown to be estimated using the observations $\{y_k\}$ corrupted by white Gaussian noise sequence $\{z_k\}$, i.e., $z_k \stackrel{\text{i.i.d.}}{\sim} \mathcal{N}(0,1)$. 

To see the connection between the Gaussian sequence model and say the white noise model, consider an orthonormal basis $\{\phi_k\}$ (e.g., Fourier Series or wavelet series) in $L_2([0,1])$, and define
\begin{equation}\label{eq:GSM_map1}
y_k = \int_0^1 \phi_k(t) \; dY(t) = \int_0^1 \phi_k(t)\; f(t) \; dt + \epsilon \; \int_0^1 \phi_k(t) \; dW(t).
\end{equation}
where the integrals with respect to $Y$ and $W$ are stochastic integrals \cite{rogersWilliams}. Define
\[
\theta_k = \int_0^1 \phi_k(t) \; f(t) \; dt
\]
and
\[
z_k = \int_0^1 \phi_k(t) \; dW(t)
\]
to get the Gaussian sequence model \eqref{eq:GaussSeqMod}. 

Let $y=(y_1, y_2, \cdots)$ be the observation sequence, and $\hat{\theta}(y)$ be an estimator of $\theta = (\theta_1, \theta_2, \cdots)$. 
If we are interested in squared error loss as in \eqref{eq:MSE}, and $\hat{\theta}$ is in $\ell_2(\mathbb{N})$, then 
\begin{equation}\label{eq:MSEGaussMod}
\begin{split}
\text{MSE}(\hat{f}) &= \Expect_f \left[\int_0^1 (f(t) - \hat{f}(Y; t))^2 \; dt \right] \\
&=  \Expect_\theta \left[ \| \theta - \hat{\theta} \|_2^2 \right] = \text{MSE}(\hat{\theta}),
\end{split}
\end{equation}
provided $\hat{\theta}$ is the transform of $\hat{f}$, i.e., 
\[
\hat{\theta}_k = \int_0^1 \phi_k(t) \hat{f}(t) dt.
\]
Thus, function estimation in the white noise model is equivalent to sequence estimation in Gaussian sequence model. 

For the regression problem \eqref{eq:RegressionMod}, mapping to a finite Gaussian sequence model 
\begin{equation}\label{eq:FiniteGaussSeqMod}
y_k = \theta_k + \epsilon \; z_k, \quad k \in \{0, 1, \cdots, N-1\}.
\end{equation} 
can be similarly defined using discrete Fourier transform. Another option is to consider discrete approximation to Fourier series. 
In both the cases, we need to take $N \to \infty$ to recover the exact infinite Gaussian sequence model \eqref{eq:GaussSeqMod}. 

The other problems, e.g., density estimation and spectrum estimation, can also be mapped to
Gaussian sequence model by appropriate mappings. See \cite{Johnstone2015Book} and \cite{tsybakov2009introduction} for details.

\subsection{Pinsker's Theorem}\label{sec:PinskersThm}
The discussion in the above two sections motivates us to restrict attention to the Gaussian sequence model
\begin{equation}\label{eq:GaussSeqMod2}
y_k = \theta_k + \epsilon \; z_k, \quad k \in \mathbb{N}.
\end{equation} 

One of the most important results in Gaussian sequence models is the famous Pinsker's theorem. 
Let $\Theta$ be an ellipsoid in $\ell_2(\mathbb{N})$ such that
\[
\Theta = \Theta(a, C) = \left\{ \theta = (\theta_1, \theta_2, \cdots) : \sum_k a_k^2 \theta_k^2 \leq C        \right\},
\]
with $\{a_k\}$ positive, nondecreasing sequence and $a_k \to \infty$. 
Consider the following \textit{diagonal linear shrinkage} estimator of $\theta$, $\hat{\theta^*}=(\hat{\theta^*_1}, \hat{\theta^*_2}, \cdots)$:
\begin{equation}\label{eq:PinskersEst}
\hat{\theta}^*_k = \left(1 - \frac{a_k}{\mu} \right)_{+} y_k. 
\end{equation}
Here, $\mu > 0$ is a constant that is a function of the noise level $\epsilon$ and the ellipsoid parameters $\{a_k\}$ and $C$; see \cite{Johnstone2015Book}. This estimator shrinks the observations $y_k$ by an amount $\left(1 - \frac{a_k}{\mu} \right)$ if 
$\frac{a_k}{\mu} < 1$, otherwise resets the observation to zero. It is linear because the amount of shrinkage used is not a function of 
the observations, and is defined beforehand. It is called diagonal, because the estimate of $\theta_k$ depends only on $y_k$, 
and not on other $y_j$, $j \neq k$.

\medskip
\begin{theorem}[Pinsker's Theorem, \cite{Johnstone2015Book}]
For the Gaussian sequence model \eqref{eq:GaussSeqMod}, the linear diagonal estimator $\hat{\theta}^*$ \eqref{eq:PinskersEst} is consistent, i.e., 
\[
\sup_{\theta \in \Theta(a, C)} \Expect_\theta \left[ \| \theta - \hat{\theta^*} \|_2^2 \right] \to 0 \quad \mbox{ as } \epsilon \to 0.
\]
Also, $\hat{\theta}^*$ is optimal among all linear estimators for each $\epsilon$. Moreover, it is asymptotically minimax over the ellipsoid $\Theta(a, C)$, i.e., 
\[
\sup_{\theta \in \Theta(a, C)} \Expect_\theta \left[ \| \theta - \hat{\theta^*} \|_2^2 \right] \sim \inf_{\hat{\theta}} \sup_{\theta \in \Theta(a, C)} \Expect_\theta \left[ \| \theta - \hat{\theta} \|_2^2 \right] \mbox{ as } \epsilon \to 0,
\]
provided condition (5.17) in \cite{Johnstone2015Book} is satisfied. 
\end{theorem}

 \medskip
We make the following remarks on the implications of the theorem.

\begin{enumerate}
 \item 
 
The Pinsker's theorem states that a linear diagonal estimator of the form 
\begin{equation}\label{eq:diagLinearEst}
\hat{\theta}_k =  c_k y_k, \quad k \in \mathbb{N},
\end{equation}
is as good as any estimator, non-diagonal or nonlinear. 

\item 
Define a Sobolev class of functions 
\begin{equation}\label{eq:SobolevFnCls}
\begin{split}
 \Sigma(\alpha, C) &= \left\{f \in L_2[0,1]: f^{(\alpha-1)} \mbox{ is absolutely } \right. \\
 & \quad \left. \mbox{ continuous and } \int_0^1 [f^{(\alpha)}(t)]^2 dt \leq C^2 \right\}.
 \end{split}
\end{equation}
 It can be shown  that a function is in the Sobolev function class $\Sigma(\alpha, \pi^\alpha C)$
 if and only if its Fourier series coefficients $\{\theta_k\}$ are in an ellipsoid $\Theta(a, C)$ with 
\begin{equation}\label{eq:FSak}
a_1 = 0, \quad \quad a_{2k} = a_{2k+1} = (2k)^\alpha. 
\end{equation}
Thus, finding the minimax function estimator over $\Sigma(\alpha, \pi^\alpha C)$ is equivalent to finding the minimax estimator 
over $\Theta(a, C)$ in the Gaussian sequence model. 
Thus, to obtain the minimax function estimator, one can take the Fourier series of the observation $Y$, shrink the coefficients as in Pinsker's theorem, and then reconstruct a signal using the modified Fourier series coefficients. See Lemma 3.3 in \cite{Johnstone2015Book}  and the discussion surrounding it for further details. 

\item 
Since $\{a_k\}$ is an increasing sequence, the optimal estimator \eqref{eq:PinskersEst} has only a \textit{finite} number of non-zero components. This is the key to the LFP classification problem we are interested in this paper. 

\end{enumerate}

\subsection{Discussion on Minimax Estimators: Shrinkage and Thresholding}\label{sec:shrinkageThresholding}

\medskip
The parameters $\theta$ need not be the Fourier coefficients of the function to be estimated. One can also take wavelet coefficients, and the Pinsker's theorem is valid for them as well. The shrinkage and reconstruction procedure has to be appropriately adjusted. 

%

\medskip
The optimal estimator is linear only because the risk is maximized over ellipsoids. For other families of $\Theta$s, the resulting optimal estimators can be nonlinear. The most significant alternatives are the nonlinear \textit{thresholding} based estimators. The basic idea is that we compute the Fourier or wavelet coefficients and set the coefficient below a threshold to zero. Thresholding based estimators are optimal when 
the elements of $\Theta$ are sparse. Threshold-based estimators and its applications to LFP classification will be discussed in detail in a future version of this paper.

\subsection{Adaptive Minimaxity: Blockwise James-Stein Estimator}\label{sec:BJS}
A major drawback of the Pinsker estimator \eqref{eq:PinskersEst} is that the optimal estimator depends on the 
ellipsoid parameters. In practice, the smoothness parameter $\alpha$, constant $C$, noise level $\epsilon$, and hence the parameter $\mu$, are not known. Thus, one does not know how many Fourier coefficients to compute, and the amount of shrinkage to use. 
Motivated by our discussion in Section~\ref{sec:ClassUsingMinimaxEst}, if we use Pinsker's estimator as a feature for learning 
a classifier, then one has to optimize over these choices of parameters using some error estimation techniques like 
cross-validation (see Section~\ref{sec:ClassifyGSM}). 
If the number of training samples is small, such an optimization can lead to overfitting. We now discuss another estimator that 
does not need the information on ellipsoid parameters, and that is also adaptively minimax over any possible choice of ellipsoid parameters. 

Let $y =(y_1, y_2, \cdots, y_n) \sim \mathcal{N}_n(\theta, \epsilon^2 \; I)$, where $\theta = (\theta_1, \cdots, \theta_n)$, be a multivariate diagonal Gaussian random vector. The James-Stein estimator of $\theta$, defined for $n > 2$, is given by
\begin{equation}\label{eq:JSfiniteD}
\hat{\theta}^{JS}(y) = \left( 1 - \frac{(n-2)\; \epsilon^2 }{\|y\|_2^2}\right)_{+} y.
\end{equation}
Note that this estimator is non-linear, as the amount of shrinkage used depends on the norm of the observation. 
Also note that here $\hat{\theta}^{JS}$ is the estimate of the entire vector $\theta$, and is a $n$ dimensional vector. The shrinkage $\left( 1 - \frac{(n-2)\; \epsilon^2 }{\|y\|_2^2}\right)_{+}$ is applied coordinate wise to each observation in the vector. 
It is well known that the James-Stein estimator is uniformly better than the maximum likelihood estimator $y$ for this problem. 

Now consider the infinite Gaussian sequence model \eqref{eq:GaussSeqMod}. A blockwise James-Stein estimator for the Gaussian sequence model is defined by dividing the infinite observation sequence $\{y_k\}$ in blocks, and by applying James-Stein estimator \eqref{eq:JSfiniteD} to each block. To define things more precisely, we need some notations. 

Consider the partition of positive integers
\[
\mathbb{N} = \cup_{j=0}^\infty \; B_j,
\]
where $B_j$ are disjoint dyadic blocks
\[
B_j = \{2^j, \cdots, 2^{j+1}-1\}.
\]
Thus, size of block $j$ is $| B_j| = 2^j$. Now define the observation for the $j$th block as
\[
y^{(j)} = \{y_i: i \in B_j\}. 
\]
Also, fix integers $L$ and $J$. The blockwise James-Stein (BJS) estimator is defined as ($j$ is used as a block index)
\begin{equation}\label{eq:BJSEst}
  \hat{\theta}^{BJS}_j=\left\{
  \begin{array}{@{}ll@{}}
    y^{(j)}, & \text{ if }\ j \leq L \\
    \hat{\theta}^{JS}(y^{(j)}), & \text{ if } L < j \leq J\\
    0& \text{ if } j \geq J,
  \end{array}\right.
\end{equation} 
where, as in \eqref{eq:JSfiniteD},
\begin{equation}
\hat{\theta}^{JS}(y^{(j)}) = \left( 1 - \frac{(2^j-2)\; \epsilon^2 }{\|y^{(j)}\|_2^2}\right)_{+} y^{(j)}.
\end{equation}
Thus, the BJS estimator leaves the first $L$ blocks untouched, applies James-Stein shrinkage to the next $J-L$ blocks, and shrinks 
the rest of the observations to zero. The integer $J$ is typically chosen to be of the order of $\log \epsilon^{-2}$. Also, 
for regression problems $\epsilon \sim \frac{1}{\sqrt{N}}$; see \eqref{eq:RegressionMod}. Thus, the only free parameter is $L$. See Section~\ref{sec:ClassifyGSM} for more details. 

Consider the dyadic Sobolev ellipsoid
\[
\Theta^\alpha_D(C) = \left\{ \theta = (\theta_1, \theta_2, \cdots) : \sum_{j\geq 0} 2^{2j\alpha} \sum_{\ell\in B_j} \theta_\ell^2 \leq C^2        \right\},
\]
and let 
\[
\mathcal{T}_{D,2} = \{ \Theta^\alpha_D(C): \; \alpha, C>0\}
\]
be the class of all such dyadic ellipsoids. The following theorem states the adaptive minimaxity of BJS estimator over $\mathcal{T}_{D,2}$.
\begin{theorem}[~\cite{Johnstone2015Book}]
For the Gaussian sequence model \eqref{eq:GaussSeqMod}, with any fixed $L$ and $J=\log \epsilon^{-2}$, the BJS estimator satisfies for each $\Theta \in \mathcal{T}_{D,2}$ 
\begin{equation}
\sup_{\theta \in \Theta} \Expect_\theta \left[ \| \theta - \hat{\theta}^{BJS} \|_2^2 \right] \sim \inf_{\hat{\theta}} \sup_{\theta \in \Theta} 
\Expect_\theta \left[ \| \theta - \hat{\theta} \|_2^2 \right] \quad \mbox{ as } \epsilon \to 0. 
\end{equation}
\end{theorem}
Thus, the BJS estimator adapts to the rate of the minimax estimator for each class in $\mathcal{T}_{D,2}$ without using the parameters of the classes in its design.

\section{Classification Algorithms based on Gaussian Sequence Model}\label{sec:ClassifyGSM}
In this section, we propose classification procedures based on the ideas discussed in previous sections. We propose two classifiers: 
the Pinsker's classifier (Section~\ref{ssec:PinskerClass}), and the Blockwise James-Stein Classifier (Section~\ref{ssec:BJSClass}). Numerical results are reported in the next section. 
\medskip

Recall from Section~\ref{sec:NPFramework} that 
the sampled LFP waveform was modeled by a regression model
\begin{equation}\label{eq:RegressLFPMod2}
Y_\ell = f( \ell / N ) +  Z_\ell, \quad \ell \in \{0, \cdots, N-1\},
\end{equation}
where $Z_\ell \stackrel{\text{i.i.d.}}{\sim}  \mathcal{N}(0,1)$, and $f \in \mathcal{F}$, where $\mathcal{F}$ is a class of smooth functions. 
Also, recall our classification problem 
\begin{equation}\label{eq:classProb2}
\begin{split}
H_k: \quad f \in \mathcal{F}_k, \quad & k \in \{1, 2, \cdots, 8\}, \\
& \quad \mathcal{F}_k \subset \mathcal{F}, \quad \mathcal{F}_k \cap \mathcal{F}_j = \emptyset, \mbox{ for } j \neq k.
\end{split}
\end{equation}      

\subsection{Pinsker Classifier}\label{ssec:PinskerClass}

We assume that the function class is a Sololev class defined in \eqref{eq:SobolevFnCls}
\[
\mathcal{F} = \Sigma(\alpha, C) ,
\]
with $\alpha$ and $C$ unknown. The classification subclasses $\{\mathcal{F}_k\}$ are to be learned from multiple samples of 
LFP waveforms. As was motivated in Section~\ref{sec:ClassUsingMinimaxEst}, we use minimax estimator as a feature. 
Pinsker's theorem allows us to choose a finite dimensional representation for the minimax estimator: use the Fourier series estimates
\begin{equation}\label{eq:FSsample}
\begin{split}
y_k = \frac{1}{N} \sum_{\ell=0}^{N-1} Y_\ell \; \phi_k ( \ell/N), \quad k \in \{1,2, \cdots, T\},
\end{split}
\end{equation}
where $\{\phi_k\}$ are the trigonometric series
\begin{equation}\label{eq:FSsinecos}
\begin{split}
\phi_1(x) &\equiv 1 \\
\phi_{2k}(x) &= \sqrt{2} \cos( 2 \pi k x) \\
\phi_{2k+1}(x) &= \sqrt{2} \sin( 2 \pi k x), \quad k=\{1, 2, \cdots\},
\end{split}
\end{equation}
and $T$ is a design parameter. Further, let $\{c_k\}$ be a sequence of shrinkage values to be selected with $c_k \in [0,1]$. Then 
our representation for the minimax estimator is
\[
\hat{\theta}_k = c_k \; y_k, \quad k = \{1, 2, \cdots, T\}.
\]
Note that we need not select $T$ and $\{c_k\}$ as in Pinsker's theorem because we do not have precise knowledge of $\mathcal{F} = \Sigma(\alpha, L)$. Moreover, since the classes $\{\mathcal{F}_k\}$ and how they are separated are also not known, it is not clear if 
the shrinkage mandated by the Pinsker's theorem is optimal for classification. 
We thus search over all possible shrinkage values (allowing for a low pass as well as a band pass filtering) in order to design a classifier. 
Thus, given a sampled LFP waveform $(Y_0, Y_1, \cdots, Y_{N-1})$ we map it to the vector $(c_1 y_1, c_2 y_2, \cdots, c_T y_T)$,
where $\{c_k\}_{k=1}^T$ and $T$ are now design parameters. 

\medskip
Once $T$ and $\{c_k\}_{k=1}^T$ has been chosen, the vector $(c_1 y_1, c_2 y_2, \cdots, c_T y_T)$ is multivariate normal random vector. Since a multivariate normal population is completely specified by its mean and covariance, one can use the training data 
to learn means and covariances of samples from different classes, and train a classifier using these learned means and covariances. 
For example, if we make the following assumptions: 
\begin{enumerate}
\item We assume that the function classes $\{\mathcal{F}_k\}$ are such that the means of the $T$-length random vector $(c_1 y_1, c_2 y_2, \cdots, c_T y_T)$ above is approximately the same for all functions within each class and differ significantly across classes. 
\medskip
\item We also assume that the covariance of $(c_1 y_1, c_2 y_2, \cdots, c_T y_T)$ is approximately the same for all functions in all the classes. 
\end{enumerate}
Then it is clear that using linear discriminant analysis (LDA) for classification would be optimal. For the LFP data we study in this paper, LDA indeed provides the best classification performance. 

\medskip
The Pinsker classifier is summarized below: 
\begin{enumerate}
\item \textit{Fix parameters}: Fix $T$ and $c_1, \cdots, c_T$.
\item \textit{Compute FS coefficients}: Use the LFP time-series to compute $T$ Fourier series coefficients $y_k$ as in \eqref{eq:FSsample}.
\item \textit{Use Shrinkage}:  Shrink the FS coefficients by $\{c_k\}$ to get feature vector $(c_1 y_1, c_2 y_2, \cdots, c_T y_T)$.
\item \textit{Dimensionality Reduction (optional)}: If LFP data is collected using multiple channels, then use principal component analysis to project the FS coefficients from all the channels (vectorized into a single high-dimensional vector) into a low dimensional subspace. 
\item \textit{Train LDA}: Train an LDA classifier.
\item \textit{Cross-validation}: Estimate the generalization error using cross-validation. 
\item \textit{Optimize free parameters}: Optimize over the choice of $L$ and  $c_1, \cdots, c_T$ fixed in step 1. 
\end{enumerate}

\subsection{Blockwise James-Stein Classifier}\label{ssec:BJSClass}
The BJS classifier can be similarly defined. 
\begin{enumerate}
\item \textit{Fix parameters}: Fix $L$ to say $L=2$. Fix $J=\log \epsilon^{-2}= \log N$. Thus, $\epsilon = 1/\sqrt{N}$. This relationship between $\epsilon$ and $N$ comes from an equivalence map between the white noise model \eqref{eq:WhiteNoiseMod} and 
the regression model \eqref{eq:RegressionMod} (see \cite{Johnstone2015Book} and \cite{tsybakov2009introduction} for more details). 
\item \textit{Compute FS coefficients}: Use the LFP time-series to compute $N$ Fourier series coefficients $y_k$ as in \eqref{eq:FSsample}. Thus, the number of FS coefficients computed is equal to the sample size $N$. 
\item \textit{Use Shrinkage}:  Use BJS shrinkage to get the feature vector $\hat{\theta}^{BJS}$
\begin{equation}
  \hat{\theta}^{BJS}_j=\left\{
  \begin{array}{@{}ll@{}}
    y^{(j)}, & \text{ if }\ j \leq 2 \\
    \hat{\theta}^{JS}(y^{(j)}), & \text{ if } 2 < j \leq \log N\\
    0& \text{ if } j \geq \log N,
  \end{array}\right.
\end{equation} 
where, as in \eqref{eq:JSfiniteD},
\begin{equation}
\hat{\theta}^{JS}(y^{(j)}) = \left( 1 - \frac{(2^j-2)}{\|y^{(j)}\|_2^2\; N}\right)_{+} y^{(j)}.
\end{equation}
\item \textit{Dimensionality Reduction (optional)}: If LFP data is collected using multiple channels, then use principal component analysis to project the FS coefficients from all the channels (vectorized into a single high-dimensional vector) into a low dimensional subspace. 
\item \textit{Train LDA}: Train an LDA classifier.
\item \textit{Cross-validation}: Estimate the generalization error using cross-validation. 
\end{enumerate}
Note that in this classifier no optimization step is needed.

\section{Classification Algorithms Applied to LFP Data}\label{sec:Numerical}

We collected $736$ samples of LFP data from $32$ channels (each channel corresponds to a different electrode) across $9$ recording sessions. All the samples had the same depth profile. This means that the $32$ electrodes were at different depths. But, those $32$-length vectors were the same for all the $736$ samples. Each sample is a data matrix of $32$ rows (corresponding to $32$ channels) and $N$ columns. The integer $N$ 
is smaller than the length of the memory period, and was a design parameter for us.  

\begin{figure}[ht]
	\center
	\includegraphics[width=8cm, height=5cm]{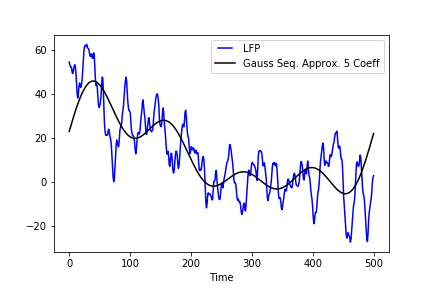}
		\hspace{0.5cm}
	\includegraphics[width=8cm, height=5cm]{LFP_10.png}
	\caption{Sample LFP waveform and its reconstruction using first 5 and 10 Fourier coefficients. Estimation is better with $10$ coefficients, but classification performance was better with $5$ coefficients. }
		\label{fig:LFPFSandRecons}
\end{figure}

For Pinsker classifier (see Section~\ref{ssec:PinskerClass}), we mapped each row ($N$ length data from each channel) to a $T$ length Fourier series coefficients vector, and used shrinkage on the Fourier coefficients by $c_k$ to get $\hat{\theta}_k$. The $32$ sets of Fourier coefficients were then appended to form a single vector of length $32 \times (2T+1)$ ($T$ sines and $T$ cosines, and one mean coefficient). A linear discriminant analysis was performed on the data after projecting the data onto $P$ principal modes. 
All the free parameters were optimized using leave-one-session-out cross-validation. 
The optimal choice of parameters was found to be:
\begin{equation}
\begin{split}
N=500, &\quad T=5, \\
& c_k=1,  \; k \leq 5, \text{ and } c_k=0, \text{ otherwise},  \mbox{ and } P=165.
\end{split}
\end{equation}
Thus, for the LFP data we analyzed, the best classification performance, of $88$\%, was obtained by low pass filtering the signal. In general, 
these choices of the free parameters would depend on the data, and may not always correspond to low pass filtering. 
In Fig.~\ref{fig:LFPFSandRecons}, we have plotted a sample LFP waveform and its reconstructions using $5$ and $10$ Fourier coefficients. Clearly, using $10$ Fourier coefficients is good for estimation, but the optimal performance for classification was found using $5$ 
coefficients. 

For the BJS classifier (see Section~\ref{ssec:BJSClass}), the process was similar to the Pinsker classifier, except the shrinkage was different and fixed. The parameter $N$ was fixed to $500$, and optimal $P$ was found to be $P=190$. The BJS classifier achieved a performance of $85$\% but had few free parameters. Due to the adaptive minimaxity properties of the BJS estimator and few free parameters, we can expect robust performance, and hence better generalization performance, of the BJS classifier. 

\begin{figure}[ht]
	\center
	\includegraphics[width=8cm, height=5cm]{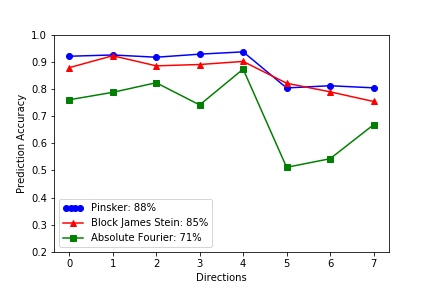}
	\includegraphics[width=8cm, height=5cm]{TargetLabels.pdf}
		\caption{Classification accuracy (conditional probability of correct decoding) per target. The average accuracy is $88$\% for the Pinsker classifier and is $85$\% for the BJS classifier.}
		\label{fig:DecodingPerformance}
\end{figure}

In Fig.~\ref{fig:DecodingPerformance}, we have plotted target wise decoding performances of the Pinsker classifier and the BJS classifier. The horizontal axis labels from $0$ to $7$ correspond to eight different target locations, also shown in Fig.~\ref{fig:DecodingPerformance}. Both the classifiers have better decoding accuracy for the targets on the left. This phenomenon was also reported in \cite{markowitz2011optimizing}. 

Also shown in the figure is the performance of another classifier, 
where the absolute values of the Fourier series coefficients were used in the Pinsker's classifier. This causes a drop in performance to $71$\%. The decoding accuracy of the targets on the right also drops by a significant margin. This suggests that phase information is crucial to this decoding task, at least for the targets on the right. We note that power spectrum based techniques, popular in the literature, use absolute values of the Fourier coefficients to compute power spectrum estimates.


\section{Conclusions and Future Work}

We proposed a new framework for robust classification of local field potentials and used it to obtain Pinsker's and blockwise James-Stein classifiers. These classifiers were then used to obtain high decoding performance on the LFP data. For future work, we will consider the following directions:

\begin{enumerate}
\item \textit{Testing with more data}: The data chosen for the above numerical results were for a particular electrode depths and were collected from a single monkey. Also, the data we used were a sub-sampled version of the broadband data collected in the experiment. We plan to test the performance of the proposed algorithm on different data sets. We expect that the optimal choice of $\{c_k\}$ and $T$, 
the number of Fourier coefficients to choose and the amount of shrinkage to use, could be different. 
\medskip
\item \textit{Apply similar techniques to Spike Rate data}: A spike rate data can also be modeled as a nonparametric function, provided the window for calculating the rate is sufficiently large, and the window is slid by a small amount. We plan to test classification performance based on spike data. Alternative ways to accommodate spike data for decoding will also be explored. 
\medskip
\item \textit{Apply Wavelet thresholding techniques}: If the function classes are not modeled using an ellipsoid, then the minimax estimator for the class might have a different structure than suggested by Pinsker's theorem. For example, if the function class $\mathcal{F}$ has a sparse representation in an orthogonal basis (like wavelets), then under certain conditions, a nonlinear thresholding 
based estimator is minimax. In another article, we will study wavelet transform based classifiers 
that are robust over a larger class of functions than the Sobolev classes. 
\end{enumerate}

\bibliographystyle{ieeetr}



\bibliography{QCD_verSubmitted}

\end{document}